\documentclass[12pt]{iopart}

\newcommand{\fourier}[1]{\mathcal{F} \biggl[#1 \biggl]}
\newcommand{\ifourier}[1]{\mathcal{F}^{-1} \biggl[#1 \biggl]}
\newcommand{\sfourier}[1]{\mathcal{F} \bigl[#1 \bigr]}
\newcommand{\sifourier}[1]{\mathcal{F}^{-1} \bigl[#1 \bigr]}

\usepackage{iopams}
\usepackage{graphicx}
\usepackage{url}

\begin{document}

\title[Handheld and low-cost digital holographic microscopy]{}

\author{Atsushi Shiraki$^{1}$, Yusuke Taniguchi$^{2}$, Tomoyoshi Shimobaba$^{2}$, Nobuyuki Masuda$^{2}$,Tomoyoshi Ito$^{2}$}

\address{$^{1}$Deparment of Information and Computer Engineering, Kisarazu National College of Technology, 2-11-1, Kiyomidaihigashi, Kisarazu-shi, Chiba 292-0041, Japan}

\address{$^{2}$Graduate School of Engineering, Chiba University, 1-33, Yayoi-cho, Inage-ku, Chiba-shi, Chiba 263-8522, Japan}

\ead{shiraki@j.kisarazu.ac.jp}

\begin{abstract}
This study developed handheld and low-cost digital holographic microscopy (DHM) by adopting an in-line type hologram, a webcam, a high power RGB light emitting diode (LED), and a pinhole. It cost less than 20,000 yen (approximately 250 US dollars at 80 yen/dollar), and was approximately 120 mm $\times$ 80 mm $\times$ 55 mm in size. In addition, by adjusting the recording-distance of a hologram, the lateral resolution power at the most suitable distance was 17.5 $\mu$m. Furthermore, this DHM was developed for use in open source libraries, and is therefore low-cost and can be easily developed by anyone. In this research, it is the feature to cut down cost and size and to improve the lateral resolution power further rather than existing reports. This DHM will be a useful application in fieldwork, education, and so forth.
\end{abstract}

Keywords DHM, handheld, low-cost, web camera, high power RGB LED 
\maketitle

\section{Introduction}
Microscope technology is used in various fields due to its: (1) High resolution, (2) Extensive field of view, and (3) Three-dimensional observation.

An optical microscope can improve the resolution power using a lens with high magnification. However, the field of view becomes narrow. It is theoretically difficult to solve this problem with an optical microscope. Digital holographic microscopy (DHM) attempts to solve this problem using a principle different to that of the optical microscope\cite{01,02}. DHM introduces holographic technology into microscope technology, giving both a high resolution power and extensive field of view, for example Ref.\cite{03}. Furthermore, DHM enables observation of an object three-dimensionally.

Although the optical system of DHM is large and expensive, low-cost DHM systems have been reported \cite{04,05}. Both studies used a digital camera for consumers as the recording device of a hologram. In Ref.\cite{04}, the optical system was constructed with a digital camera made by Canon, a laser diode, and some lenses. The cost was approximately \$1,000, and it realized a lateral resolution power of 2.8 $\mu$m. The system was about 35 cm in size in the optical axis direction. Ref.\cite{04} is the first paper to suggest using a digital camera for consumers to make low-cost DHM. However, the system was not conducive to handheld size, and the cost was slightly higher. In addition, real-time reconstruction was difficult because the system used a digital camera for consumers\cite{06}.

In our study, we succeeded in reducing the cost and the size of DHM. The cost is less than 20,000 yen (approximately 240 US dollars at 83 yen/dollar), and it is approximately 120 mm $\times$ 80 mm $\times$ 55 mm in size. In addition, by adjusting the recording distance of a hologram, the lateral resolution power at the most suitable distance is 17.5 $\mu$m. Therefore, this DHM is useful in fieldwork, education, and so forth. In addition, we can obtain a reconstruction image in real-time by using a web camera and graphics processing unit (GPU) acceleration\cite{06}.

This paper is structured as follows. Section 2, describes the recording and reconstruction method of a hologram. Section 3 and 4 explain the hardware configuration and the software used. Section 5 evaluates this system. The final section discusses the study results and proposes further studies.

\section{Recording and reconstruction method}
\begin{figure}[t]
  \begin{center}
    \includegraphics[keepaspectratio=true,height=50mm]{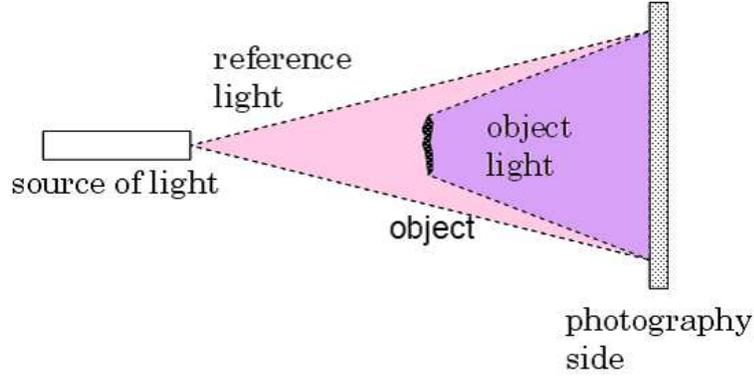}
  \end{center}
  \caption{Conceptual figure of in-line type DHM.}
  \label{fig:in-line.eps}
\end{figure}

This study aims to achieve price reduction and downsizing of DHM. Therefore, we adopted the in-line type DHM as the recording setup of holograms, which does not need optical parts, such as lenses\cite{07,08}. The conceptual figure of the in-line type DHM is shown in Figure \ref{fig:in-line.eps}. Thus, there is no need to divide the object light and reference light unlike in Mach-Zehnder type DHM, which reduces the cost of lenses. 

Next, we explain the reconstruction technique. In DHM, computer simulation performs reconstruction from a hologram. Here, in order to pursue real-time reconstruction, it is necessary to change the formula of optical propagation into a form that is easy to control by computer. The angular spectrum method, Fresnel diffraction, Fraunhofer diffraction, and so forth are approximation techniques. These are properly used according to conditions, such as propagation distance. In order to realize the downsizing of DHM, which is one of the purposes of this study, it is necessary to shorten the distance between the object and the hologram side (charge coupled device (CCD) camera). Therefore, we adopted the angular spectrum method, which is effective when the distance of the object and the hologram side is short. The angular spectrum method is shown below. 

\begin{eqnarray}
A(f_{x}, f_{y}) &=& \fourier{a(\xi, \eta) }, \nonumber\\
                &=& \int \int ^{\infty}_{-\infty} a(\xi, \eta) \exp \bigl\{-2 \pi i (\xi f_{x} + \eta f_{y}) \bigr\} d \xi d \eta.\\
U(f_{x}, f_{y}) &=& A(f_{x}, f_{y}) \exp (-2 \pi i f_{z} d).\\
u(x, y) &=& \ifourier{U(f_{x}, f_{y})}, \nonumber\\
        &=& \int \int^{\infty}_{-\infty} U(f_{x}, f_{y}) \exp \bigl\{2 \pi i (x f_{x} + y f_{y}) \bigr\} d f_{x} d f_{y}.
\end{eqnarray}

Here, the operators $\sfourier{\cdot}$ and $\sifourier{\cdot}$ indicate Fourier and inverse Fourier transforms, and the subscripts $x, y$, and $z$ indicate the horizontal, vertical, and depth components. ($\xi$, $\eta$) and ($x$, $y$) are the coordinates of the hologram side and the observation side, respectively. $d$ is the distance between the hologram side and the observation side, and $\lambda$ is the wavelength of the light. $f_{x}$, $f_{y}$ and $f_{z}$ are the spatial frequency components of each axis, and $f_{z}$ is shown as follows.

\begin{equation}
f_{z} = \sqrt{\frac{1}{\lambda^{2}}-f_{x}^{2}-f_{y}^{2}}.
\end{equation}

If we calculate using the angular spectrum method, Discrete Fourier Transform (DFT) is calculated on a large scale and at high-speed by Fast Fourier Transform (FFT). Therefore, it is possible to perform recording of a hologram and reconstruction by computer simulation, and if we use a GPU, the reconstruction is obtained in real-time.

\section{Hardware design}
\begin{figure}[t]
  \begin{center}
    \includegraphics[keepaspectratio=true,height=60mm]{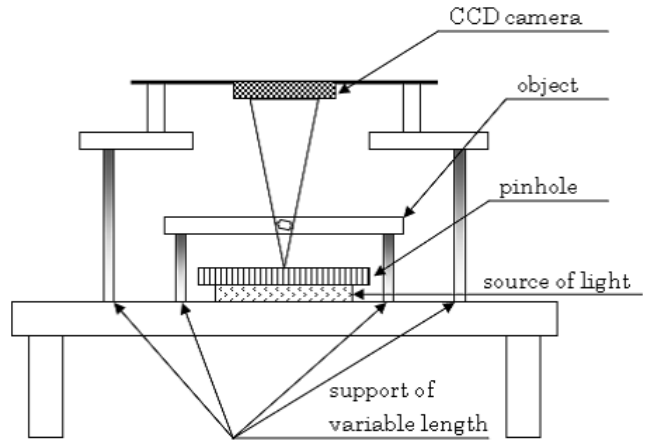}
  \end{center}
  \caption{Composition of the handheld DHM system to create.}
  \label{fig:dhm-fig.eps}
\end{figure}

In order to realize low-cost and downsizing of DHM, we build hardware as shown in Figure \ref{fig:dhm-fig.eps}. 
We used a web camera, HD Pro Webcam C910 made by Logicool, as the CCD camera and removed the lens of the web camera for focusing an image on the CCD camera because the lens is not required for DHM application. The resolution for photography of this web camera is 1920$\times$1080 pixels. Using a commercial web camera, the cost can be sharply reduced. The OSTCXBC1C1S, which is a high power RGB- light emitting Diode (LED) made by OptoSupply, is used as the light source. This LED features the ability to allow a 350 mA current to be passed through it. Another feature is having three wavelengths, 470 nm, 525 nm, and 625 nm. The light of this LED is used as the point light source passing through the pinhole. The diameter of the pinhole is 5$\mu$m (the product number of the pinhole made by KOYO is 2412). Although the pinhole is generally positioned on the focal side of an object lens, in our study, the perfect lensless optical system is used instead of an object lens for cost reduction. As a substitute, a pinhole is directly attached to the LED. 

\begin{figure}[t]
  \begin{center}
    \includegraphics[keepaspectratio=true,height=52mm]{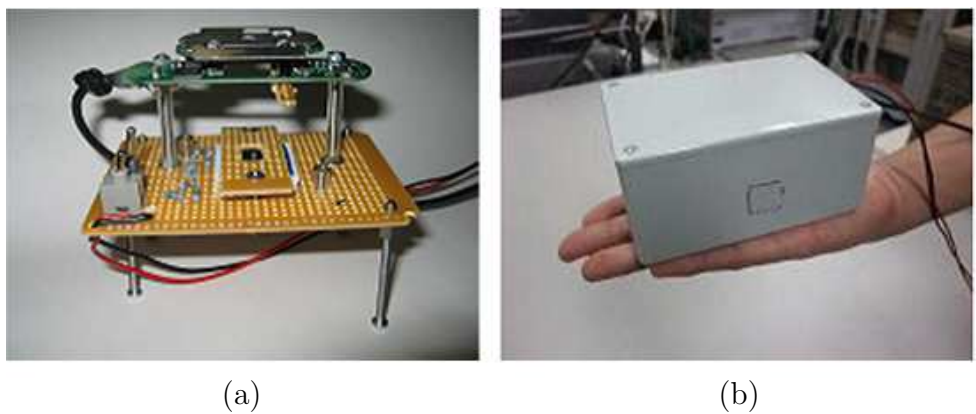}
  \end{center}
  \caption{Photograph of DHM system. (a) DHM system without a shading case. (b) Handheld DHM system.}
  \label{fig:hardware.eps}
\end{figure}

Moreover, in order to adjust the distance of the light source, an object, and a CCD camera, variable length supports are used. The actual DHM system is shown in Figure \ref{fig:hardware.eps}. 
When recording a hologram, the actual DHM system of Figure \ref{fig:hardware.eps} is housed in a shading case.

\section{Software design}
Software for recording a hologram and the reconstruction from a hologram is required. OpenCV \cite{09}, which is the open source library by which Intel is developed and released, is used to record the hologram. Using OpenCV, control of a web camera and editing of a recorded hologram can be performed. Moreover, it is necessary to calculate the diffraction of light for reconstruction from a hologram. Therefore, the CWO++ library, which is an open source library for optical calculation, is used\cite{10}. The CWO++ library is equipped with several commands used when optical calculation is performed. Therefore, calculation of the angle spectrum method adopted as this research can calculate on CPU or GPU by a simple description. Thus, use of the OpenCV and CWO++ library is cost free, enables simplification of programming and improvement in the calculation speed. 

\vspace{-11.5cm}
    \hspace{3.3cm}
        (a)
    \hspace{5.8cm}
        (b)
\vspace{11.5cm}

\section{Performance}
\subsection{Cost and size}
The costs of the parts used are shown in Table \ref{tab:cost-parts}. As shown in Table \ref{tab:cost-parts}, it is constituted for less than 20,000 yen (approximately 250 US dollars at 80 yen/dollar). Moreover, the size of DHM is equal to the size of the shading case, and it is 120 mm $\times$ 80 mm $\times$ 55 mm.

\begin{table}[t]
 \caption{Parts used and costs for the DHM system.}
 \begin{center}
  \begin{tabular}{|c|c|}
    \hline
     parts  & costs[yen]   \\
    \hline
     CCD camera  & 7,092   \\
    \hline
     high power RGB-LED  & 500   \\
    \hline
     pinhole  & 9,800   \\
    \hline
     shading case  & 1,400   \\
    \hline
     total  & 18,792   \\
    \hline
  \end{tabular}
 \end{center}
 \label{tab:cost-parts}
\end{table}
 
\subsection{Measurement of the lateral resolution power}
\begin{figure}[t]
  \begin{center}
    \includegraphics[keepaspectratio=true,height=43mm]{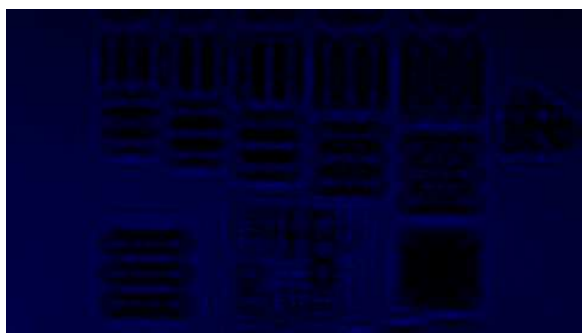}
  \end{center}
  \caption{Hologram of element number 5 of group number 4 of the USAF test target.}
  \label{fig:hologram.eps}
\end{figure}

\begin{figure}[t]
  \begin{center}
    \includegraphics[keepaspectratio=true,height=43mm]{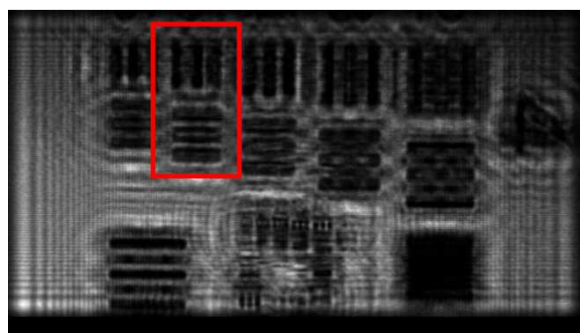}
  \end{center}
  \caption{Reconstruction image of element number 5 of group number 4 of the USAF test target.}
  \label{fig:reconstruction.eps}
\end{figure}

In order to measure the lateral resolution power of the DHM system created in this study, the USAF1951 test pattern was used. In addition, a blue light with the shortest wavelength was used as the light source. To increase the lateral resolution power, we adjusted the positions of the CCD camera and the object using the variable length support. In this system, the lateral resolution power was highest when the distance from the light source to the object was 3 mm and the distance from the light source to the CCD camera was 13 mm. The hologram recorded in this condition is shown in Figure \ref{fig:hologram.eps}. Furthermore, the reconstruction image obtained from Figure \ref{fig:hologram.eps} is shown in Figure \ref{fig:reconstruction.eps}. Figure \ref{fig:hologram.eps} and Figure \ref{fig:reconstruction.eps} show that those have resolved element number 5 of group number 4. As mentioned above, the lateral resolution power of this DHM is 17.5 $\mu$m.

\section{Conclusion and discussion}
In this study, we reduced the cost of DHM  to less than 20,000 yen (approximately 250 US dollars at 80 yen/dollar), and reduced the size to 120 mm $\times$ 80 mm $\times$ 55 mm. The lateral resolution power of this DHM was 17.5 $\mu$m. The downsizing of DHM is useful for fieldwork, and the reduction in the cost of DHM and use of an open source library are of great benefit in educational environments. Moreover, since the LED light source of RGB is used in this study, various wavelengths can be controlled. This is applicable to wavefront recovery using two-wave digital holography\cite{02}, two-wave Laplacian reconstruction\cite{11}, or two-wave Laplacian reconstruction\cite{12}. 

In this study, since the angular spectrum method was used, a reconstruction image could be obtained in the same resolution as a hologram. The scaled angular spectrum method \cite{13} that can calculate diffraction at different sampling rates on a hologram and reconstructed image is useful for this DHM system because it enables observation of a detailed reconstructed-image with a smaller sampling rate on the reconstructed image. Adding such a function of DHM and increasing convenience are future study considerations.


\end{document}